\RequirePackage{fix-cm}
\documentclass[smallextended]{svjour3}       
\smartqed  
\usepackage{graphicx}
\begin{document}

\title{Patterned Supersolids in Dipolar Bose Systems
}


\author{Youssef Kora         \and
        Massimo Boninsegni 
}


\institute{
              Department of Physics, University of Alberta, Edmonton, Alberta, Canada, T6G 2E1\\
              \email{m.boninsegni@ualberta.ca}
}

\date{Received: date / Accepted: date}

\maketitle

\begin{abstract}
We study by means of first principle Quantum Monte Carlo simulations the ground state phase diagram of a system of dipolar bosons with aligned dipole moments, and with the inclusion of a two-body  repulsive potential of varying range. The system  is shown to display a supersolid phase in a relatively broad region of the phase diagram, featuring different crystalline patterns depending on the density and on the range of the repulsive part of the interaction (scattering length). The supersolid phase is sandwiched between a classical crystal of parallel filaments and a homogeneous superfluid phase. We show that a ``roton'' minimum appears in the elementary excitation spectrum of the superfluid as the system approaches crystallization. The predictions of this study are in quantitative agreement with recent experimental results.
\end{abstract}

\section{Introduction}
\label{intro}
Dipolar gases are the focus of extensive experimental and theoretical research, motivated by the possibility that yet unobserved, exotic phases of matter may be underlain by the distinctive character of the  inter-particle interaction, both long ranged and anisotropic \cite{menotti}. In particular, the experimental achievement of Bose-Einstein Condensation of  atomic systems with large magnetic moments \cite{griesmeier,lu,aikawa,depaz,ni,yan,takekoshi,park,balewski}
suggests that one might be able to predict and observe phases featuring more than one type of order.
Of particular interest is the supersolid phase, which spontaneously breaks both continuous translational and global U(1) symmetries, thus featuring simultaneously crystalline order and flow without dissipation (see, for instance, Ref. \cite{rmp}). 
\\ \indent
As supersolid behavior in a crystal of $^4$He, once deemed the most promising candidate, has so far eluded unambiguous observation, dipolar bosons have been suggested as a likely physical setting for its detection in various experimental \cite{wentzel,ferlaino} and theoretical \cite{cinti10,jltp,cinti} works. In Ref. \cite{cinti}, for instance,  it is contended  that, if dipole moments are aligned, an ordered array of filaments (or, prolate droplets) constitutes the ground state of the system in rather broad conditions, and preliminary evidence of global phase coherence among such droplets was offered for specific values of density and interaction parameters.
\\ \indent 
It seems now accepted that the appearance of filaments in this system is the result of the competition between the attractive part of the dipolar interaction, and the presence of a short range repulsion, which prevents the system from collapsing. Such a repulsive part is often modeled theoretically through the so-called scattering length approximation. Because the scattering length is experimentally controllable and can be varied  by means of the Feshbach resonance (see, for instance, Ref. \cite{feshbach}),  the possibility arises of exploring the quantum phase diagram of the system in its entirety, at least within known limitations (e.g., three-body recombination) \cite{fb,pfau2,ferlaino3,ferlaino2}. A comprehensive theoretical study of the bulk phase diagram of the system at low temperature based on reliable, first principle computational methods, aimed at helping in the design and interpretation of present and future experiments, seems therefore timely \cite{lastnote}. 
\\ \indent 
In this work, we carry out state-of-the-art Quantum Monte Carlo (QMC) simulations to investigate the low temperature phase diagram of dipolar bosons with aligned  dipole moments. We model the repulsive part of the interaction by means of an inverse power law potential, as done in previous work \cite{cinti}; a straightforward connection exists   between the characteristic range $\sigma$ of this interaction, and the scattering length $a_s$. We map out the phase diagram  by computing relevant correlations, as well as the superfluid density, all directly accessible in our numerical approach, as a function of particle density and $\sigma$.
\\ \indent 
The system displays several distinct phases, ranging from an essentially classical crystal of parallel, particle-thin filaments in one limit ($\sigma\to 0$), to a hard-sphere-like superfluid, reminiscent of liquid $^4$He in the opposite ($\sigma\to\infty$) limit, at any particle density. An intermediate region exists between these two phases, in which the ground state displays both crystalline order and a finite superfluid response. This {\em supersolid} phase exists within a range of scattering length that  depends monotonically on the density. 
\\ \indent
The crystal structure of the supersolid is quite distinct from that of the classical crystal,  determined by quantum-mechanical effects, both zero-point motion as well as exchanges of particles (as already noted in Ref. \cite{cinti}). Remarkably, the supersolid phase features 
different arrangements of particles (patterns) in different regions of the phase diagram. At low density, the supersolid phase consists of  a crystal of prolate droplets, with quantum-mechanical exchanges of particles across droplets, as suggested in Ref. \cite{cinti}. As the density is increased, on the other hand, the patterns resemble some of the periodic structures originally predicted to occur in two-dimensional dipolar systems \cite{vanderbilt,spivak}.
\\ \indent 
We study the elementary excitation spectrum for the superfluid phase, which features the experimentally observed ``roton'' minimum \cite{ferlaino}, as crystallization is approached from the superfluid side, {\em both} on {\em reducing} $\sigma$ (i.e., moving toward the supersolid phase) as well as on {\em increasing} it, in which case the system behaves essentially as a hard sphere fluid, transitioning into a conventional (i.e., non-superfluid) crystal.
\\ \indent
The remainder of this paper is organized as follows: in section \ref{mo} we describe the model of the system; in Sec. \ref{me} we briefly describe our methodology; we present and discuss our results in Sec. \ref{else} and finally outline our conclusions in Sec. \ref{conc}, where we discuss the relevance of this study to recent experimental work.

\section{Model}\label{mo}
The system is modeled as an ensemble of $N$ identical particles of mass $m$ and spin zero, hence obeying Bose statistics. These particles possess a magnetic moment $d$, pointing in the  $z$-direction. We are interested in studying the phase diagram of the {\em bulk}; thus, unlike in a typical experiment, we do not confine the simulated system by means of an external potential. Rather, our system is enclosed in a three dimensional box, shaped like a cuboid of volume $V$, with periodic boundary conditions in the three directions.  The shape of the cell was varied, depending on the particular structures and patterns forming at the various physical conditions. We take the characteristic length of the dipolar interaction, $a \equiv md^2/\hbar^2$ as our unit of length, and $\epsilon \equiv \hbar^2/(ma^2)$, as that of energy and temperature.  The quantum-mechanical many-body Hamiltonian in dimensionless form reads as follows:
\begin{eqnarray}\label{u}
\hat H = - \frac{1}{2} \sum_{i}\nabla^2_{i}+\sum_{i<j}U({\bf r}_i,{\bf r}_j)
\end{eqnarray}
where the first (second) sum runs over all particles (pairs of particles), and the pair potential consists of two parts, $U({\bf r},{\bf r}') = U_{sr}(|{\bf r}-{\bf r}'|) + U_d({\bf r},{\bf r}')$, $U_{sr}$ being the repulsive part. 
As mentioned above, in most theoretical studies the repulsive part of the interaction is modeled by means of the so-called scattering length approximation, namely
\begin{equation}\label{sla}
U_{sr}(r) = \frac{4\pi\hbar^2 a_s}{m} \ \delta({\bf r})
\end{equation}
To the extent that such a representation is valid, expression (\ref{sla}) can be replaced by any other potential that has the same scattering length $a_s$. In this work, we use for $U_{sr}$ 
the repulsive part of the standard Lennard-Jones potential, i.e.,
\begin{equation}\label{Usr}
U_{sr}(r)=(\sigma/r)^{12}
\end{equation}
whose use is more convenient in numerical simulations. The parameter $\sigma$ of the potential $U_{sr}$ used here is directly related to the scattering length $a_s$, through
\begin{equation}\label{as2}
\frac{a_s}{\sigma} \approx 0.76\ \sigma^{1/5}
\end{equation}
(see, for instance, Ref. \cite{flugge}).
$U_{d}$ is the classical dipolar interaction between two aligned dipole moments, namely
\begin{equation}\label{Ud}
U_d({\bf r},{\bf r}')=\frac{1}{|{\bf r}-{\bf r}'|^3}\left(1-\frac{3(z-z')^2}{|{\bf r}-{\bf r}'|^2}\right)
\end{equation}
\\ \indent 
At zero temperature there are two parameters that govern the physics of the system, namely the particle density $\rho\equiv N/V$, and the characteristic range $\sigma$ of the repulsive interaction. The classical limit is approached as $\sigma\to0$, whereupon the attractive well of the anisotropic interaction becomes deep enough to dominate the physics, and quantum mechanical effects are small. On the other hand, as $\sigma$ is increased, quantum mechanical effects are increasingly prominent, eventually destabilizing the classical ground state and giving rise to interesting physics. 

\section{Methodology}\label{me}
We carry out QMC simulations of the system described in section \ref{mo} using the continuous-space worm algorithm \cite{worm,worm2}. We shall not review the details of this method, referring instead the reader to the original references. We utilized a canonical variant of the algorithm in which the total number of particles $N$ is held constant, in order to simulate the system at fixed density \cite{mezz1,mezz2}. Although this is a finite temperature technique, we perform simulations at sufficiently low temperatures, so that computed physical properties are essentially those of the ground state.
\\ \indent 
We survey the phase diagram of the system by performing simulations at a fixed density for different values of $\sigma$, and then repeating the process for different values of the density. We performed simulations of systems of sizes ranging from $N=160$ to $N=648$ particles, and densities between $\rho=0.125$, which is close to the value of some of the current experiments \cite{ferlaino}, up to three orders of magnitude higher, i.e., $\rho=100$, quite likely not attainable in present time experiments but nonetheless of fundamental interest.
To ensure that the physics is independent of the initial configuration of the particles, most simulations were started from high-temperature, disordered configurations. 
\\ \indent
Details of the simulation are standard; we made use of the primitive approximation for the high-temperature density matrix. Although more accurate  forms exist, we found that in practice with this particular interaction the primitive approximation is the most efficient. All of the results quoted here are extrapolated to the limit of time step $\tau\to 0$; quite generally we found numerical estimates for structural and superfluid properties of interest here obtained with a value of the time step $\tau\sim 10^{-3}\epsilon^{-1}$ to be indistinguishable from the extrapolated ones, within the statistical uncertainties of the calculation.
\\ \indent
Occurrence of crystalline order in the system can be detected through the calculation of the static structure factor $S({\bf q})$; because of the anisotropy of the interaction and the ensuing tendency of the system to form filaments along $z$ (the direction of alignment of the dipole moments), we computed $S({\bf q})$ for {\bf q}-vectors lying in the $xy$ plane. As mentioned above, distinct, characteristic patterns form for different values of the parameters $\rho, \sigma$. The identification of the various patterns is achieved through visual inspection of the configurations generated by the algorithm in the course of sufficiently long computer runs.
\\ \indent
The superfluid response of the system is assessed through the direct calculation of the superfluid fraction $\rho_S(T)$, by means of the well-known winding number estimator \cite{pollock}. Due to the anisotropic nature of the interaction, and consequently of the superfluid (and supersolid) phases observed here, we offer results for the in-plane ($xy$) superfluid response only, throughout this paper. A typical result for the in-plane superfluid fraction $\rho_S$ is shown in Fig. \ref{phase}b.

\section{Results}\label{else}
\subsection{Phase diagram}\label{phasediagram}
\begin{figure}[h]
\centering
\includegraphics[width=1\textwidth]{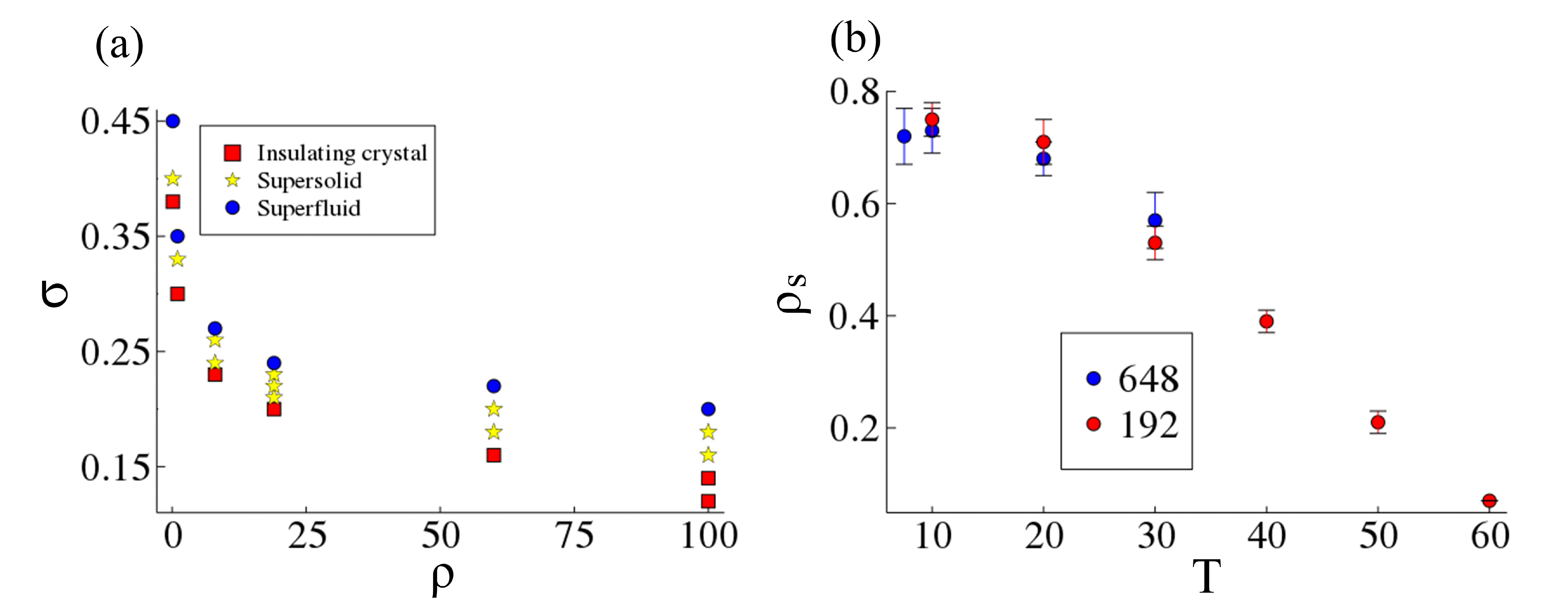}
\caption{  (a) Ground state phase diagram of the system in the $(\rho,\sigma)$ plane. Each phase is determined through simulations as explained in the text. (b) Superfluid fraction ($xy$ plane) of the system at $\rho=100$ and $\sigma=0.18$ as a function of temperature, and for two system sizes.}
\label{phase}
\end{figure}

\begin{figure}[h]
\centering
\includegraphics[width=0.85\textwidth]{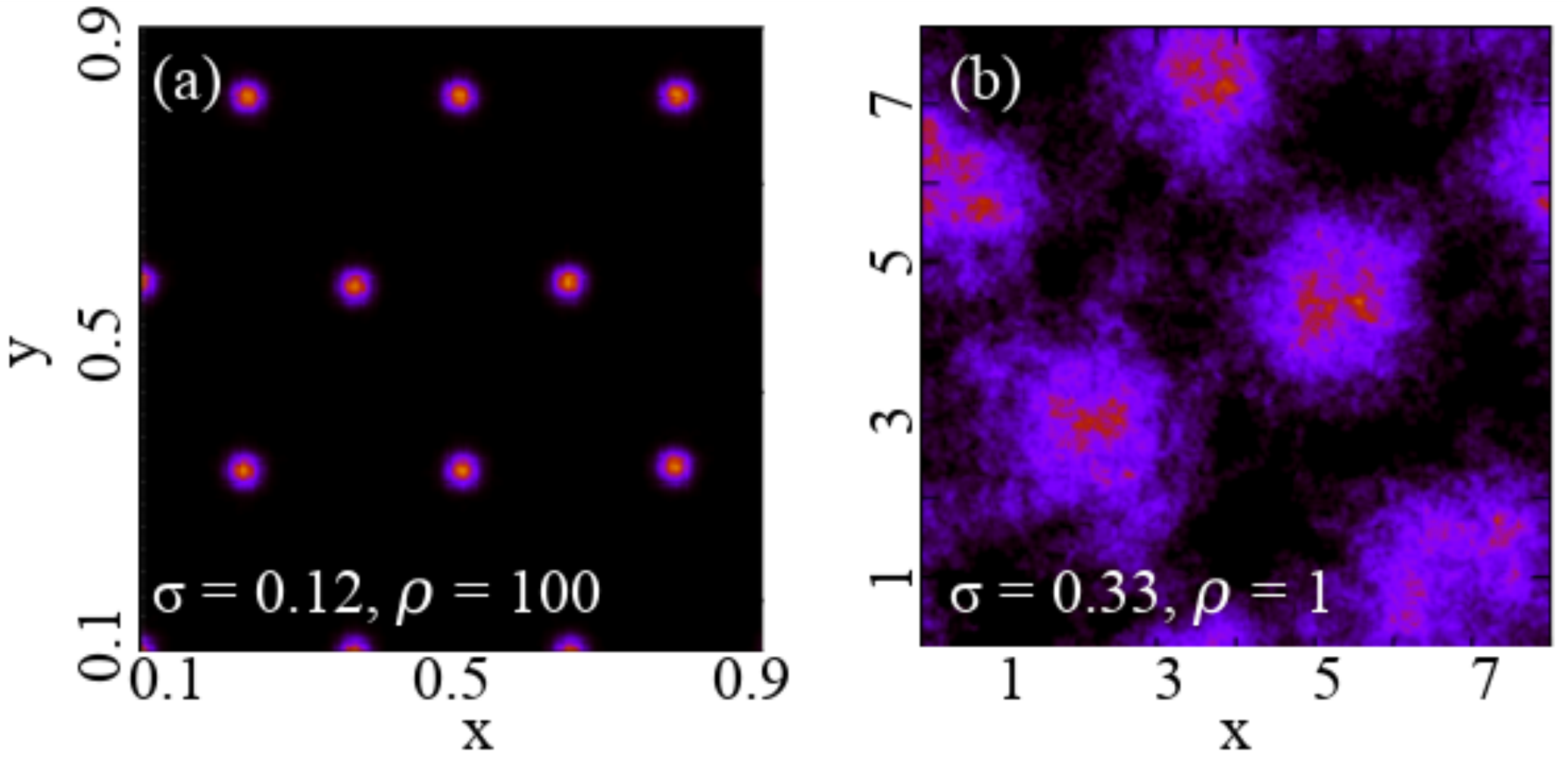} 
\caption{ Density maps of the system at density $\rho=100$ for $\sigma=0.12$ (a), and  $\rho=1$ for $\sigma=0.33$ (b), integrated along the $z$-direction (i.e., the direction of dipole moment alignment). Brighter areas indicate higher density. The maps are obtained from particle world lines of a single configuration. In (a), the physics of the system is dominated by the potential energy, and the ground state is essentially the classical one, consisting of an array of particle-thin, parallel filaments. In (b), the ground state is a crystal of droplets with frequent quantum mechanical exchanges among adjacent droplets, leading to a finite superfluid response (as such displaying supersolid behavior).}
\label{str1}
\end{figure}

The ground state phase diagram of the system, as it emerges from our extensive QMC simulations, is shown in Fig. \ref{phase}a in the $\rho-\sigma$ plane. Here we outline some of its generic features, offering a more detailed discussion of the two most interesting phases (the supersolid and the superfluid) in subsections \ref{supersolid} and \ref{superfluid}. 
\\ \indent
For any fixed value of the density, two clear physical limits can be easily identified, more intuitively discussed in terms of the average interparticle distance $b\equiv \rho^{-1/3}$. Specifically, if $\sigma << b$ the ground state of the system consists of an ordered array (a triangular lattice) of thin filaments oriented along the $z$-direction. An example is shown in Fig. \ref{str1}a, displaying the density of the system, integrated along the  $z$-direction \cite{morepr}. 
In this physical limit, the depth of the attractive well of the dipolar interaction causes the potential energy to dominate  the behavior of the system, which can be understood and predicted quantitatively along classical lines, as shown in Ref. \cite{cinti}. Specifically, the system forms an ordered array of particle-thin parallel filaments, arranged on a triangular lattice. In this case, exchanges of identical particles are suppressed, both within a filament, as well as across filaments.
\\ \indent
Conversely, when $\sigma$ becomes of the order of the interparticle distance $b$, the attractive well of the dipolar interaction weakens, and the physics of the  system morphs into that of a hard-sphere fluid, as the repulsive part of the interaction becomes the dominant feature. The behavior of the system in this limit is very similar to that of superfluid $^4$He, which undergoes ``conventional" crystallization as $\sigma \sim b$ (this part of the phase diagram is not shown in Fig. \ref{phase}a).
\\ \indent
As shown in Fig. \ref{phase}a, there is in an intermediate range of $\sigma/b$ wherein the system displays the most interesting, novel behavior, specifically a supersolid phase, which we discuss in detail in subsection \ref{supersolid}. A remarkable feature of this phase diagram is that there is no way of going from the superfluid to the insulating crystalline phase at zero temperature, by varying $\rho$ and/or $\sigma$, without going through a supersolid phase. This is also the case in a  2D system of soft core bosons \cite{saccani}.

\indent
\subsection{Supersolid}\label{supersolid}
\begin{figure}[h]
\centering
\includegraphics[width=0.85\textwidth]{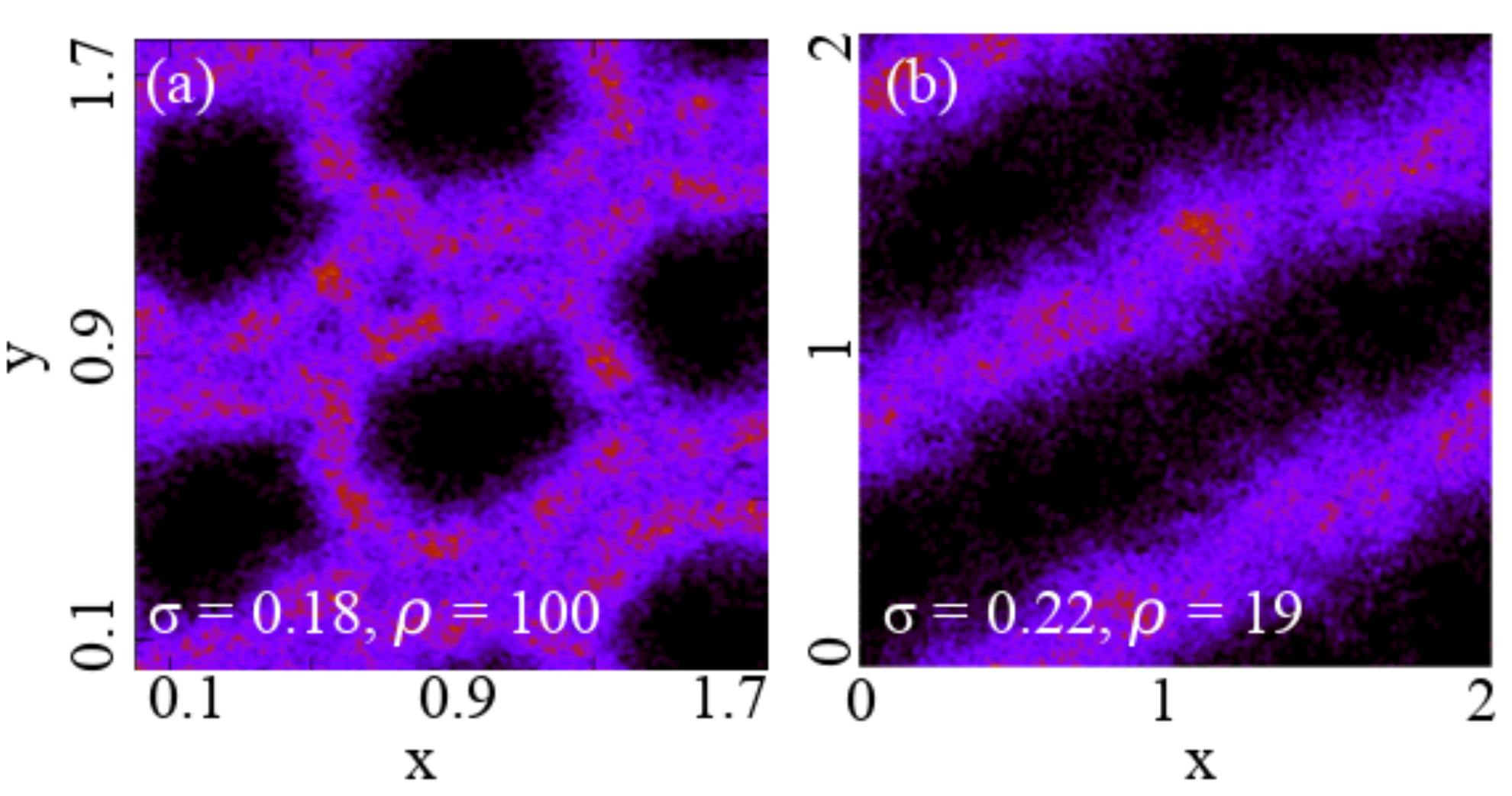} 
\caption{  Density maps of the system at density $\rho=100$, for $\sigma=0.18$ (a), and $\rho=19$, for $\sigma=0.22$ (b), integrated along the $z$-direction. Brighter areas indicate higher density. These maps are obtained from particle world lines of a single configuration.}
\label{str2}
\end{figure}

The supersolid phase is generally characterized by the formation of relatively large, prolate droplets, elongated along the $z$-direction, arranging themselves on a triangular lattice or forming more complex structures, as we discuss below. Quantum mechanical exchanges of identical particles, largely suppressed in the classical filament crystal, become important, initially within a single filament \cite{super} and progressively across filaments, establishing phase coherence and leading to a finite, three-dimensional superfluid response throughout the system. It is interesting to note that while the superfluid response is anisotropic, and in the $xy$ plane saturates to a value lower than unity as $T\to 0$ (see Fig. \ref{phase}b), as expected \cite{leggett}, it is always seen to approach unity, in the same limit, in the direction of elongation of the droplets ($z$). 
\\ \indent
The  range of values of $\sigma$ within which the system displays supersolid behavior depends on the density of the system, as shown in Fig. \ref{phase}a; in particular, at high density the supersolid occurs for lower values of $\sigma$, as quantum-mechanical exchanges are favored by a lower interparticle distance.
The width in $\sigma$ of the supersolid region, however, appears to be roughly independent of $\rho$.
\\ \indent 
As mentioned above, the occurrence of an ordered arrangement can be established through the calculation of the static structure factor, which develops (Bragg) peaks in correspondence of relevant wave vectors (e.g., the inverse distance between droplets). Additional information comes from the direct visual inspection of the patterns that form in the course of the simulation, which are quantum-mechanical in nature and can be markedly different, depending on both the density and the value of $\sigma$.
\\ \indent
Fig. \ref{str1}b shows the same type of density map as in Fig. \ref{str1}a, but for a supersolid system. As one can see, the filaments are in this case replaced by larger droplets, which include considerably more particles than the classical filaments \cite{important}. Quantum-mechanical exchanges among adjacent  droplets are frequent at low temperature (i.e., $T\sim 1$ for this particular choice of $\rho,\ \sigma$); consequently, macroscopic exchange cycles (i.e., comprising nearly {\em all} the particles in the system) take place.
Clearly, the density map shown in Fig. \ref{str1}b is very reminiscent of those that appear in a purely two-dimensional (2D) system of soft core bosons \cite{saccani,saccani2}, namely the simplest supersolid. Indeed, the physics of the system under study here, projected onto the plane perpendicular to the filaments, could be regarded as equivalent to that of a 2D soft core system, the third dimension serving the purpose of “piling up” particles, thus allowing for the formation of ``cluster'' unit cells.
\\ \indent
It is quite interesting to note that the supersolid phase does not always display the structure shown in Fig. \ref {str1}b. Rather, at high particle density patterns begin to emerge, such as the inverted droplet structure shown in Fig. \ref{str2}a, and the striped one \cite{cintiboninsegni} of Fig. \ref{str2}b, or others that are evocative of those predicted for 2D dipolar systems, in the context of ``microemulsions'' \cite{vanderbilt,spivak}

\subsection{Superfluid}\label{superfluid}
\begin{figure}[h]
\centering
\includegraphics[width=0.91\textwidth]{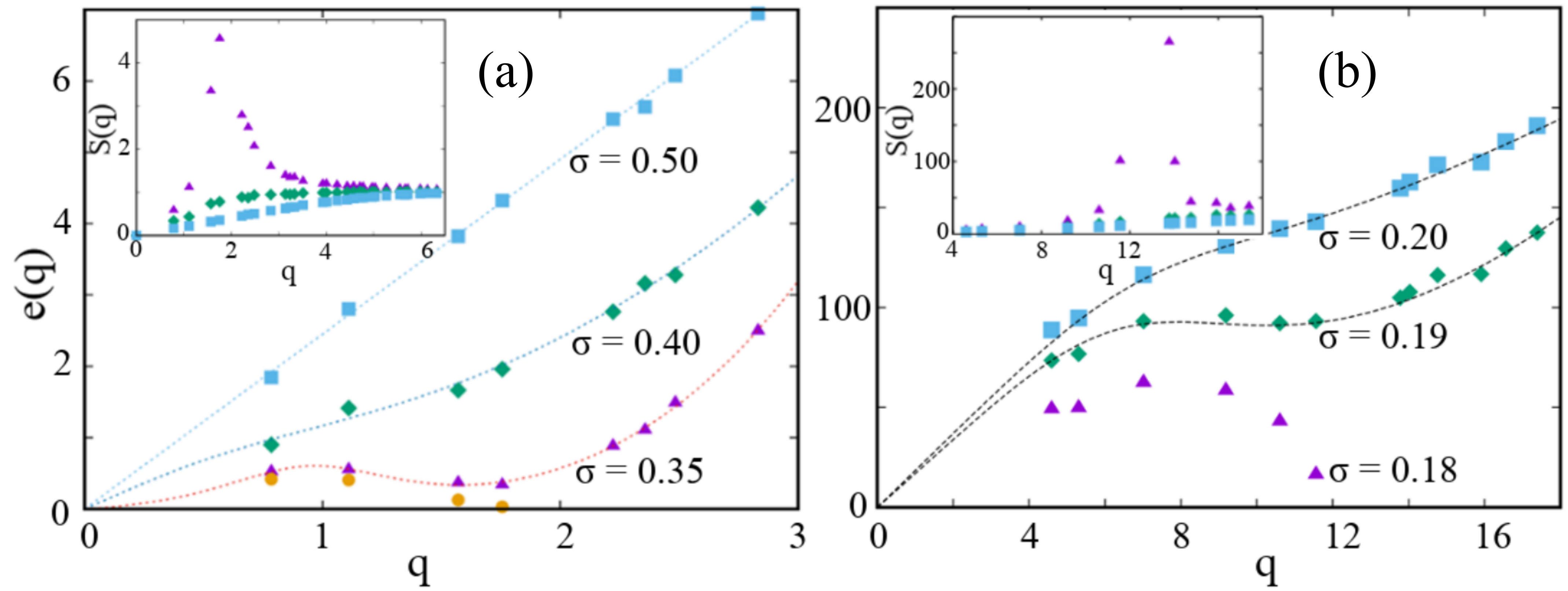}
\caption{  Elementary excitation spectra of a system of dipolar bosons for different values of the repulsive radius $\sigma$ near the superfluid-supersolid transition, computed using eq. \ref{bijl}. Results shown here are for $\rho=1$ (a) and $\rho=100$ (b), in the low temperature limit. Statistical errors are smaller than the size of the symbols. Solid  lines are fits to the data using Pad\'e approximants, and are only meant as a guide to the eye. Insets show the corresponding in-plane static structure factor $S(q)$, averaged over all directions. In (a), yellow circles refer to $\sigma=0.33$, for which crystal order appears, i.e., the system is in the supersolid phase. In (b), this happens at $\sigma=0.18$}
\label{eq}
\end{figure}

The superfluid phase arises as the range of the repulsive interaction $U_{sr}$ (eq. \ref{Usr}) is progressively increased, at fixed density, as shown in Fig. \ref{phase}a. As the droplets expand, due to the hard core repulsion, they eventually merge, giving rise to a uniform superfluid phase. It is interesting to study the elementary excitation spectrum of the superfluid phase, especially as crystallization is approached.
\\ \indent
Although it is possible to infer the excitation spectrum from the full imaginary time dynamics, computed by QMC, using an inversion method (e.g., MaxEnt \cite{ceperley,maxent}), because we are only interested here in gaining qualitative understanding we make use of a simpler approach, based on 
the Bijl-Feynman approximation \cite{feynman54}. Specifically, we assume that the dynamic structure factor $S({\bf q},\omega)$ is dominated by a single peak, which allows one to obtain the elementary excitation spectrum as
\begin{equation}\label{bijl}
e(q) = \frac {q^2}{2S(q)}
\end{equation}
Eq. \ref{bijl}, known as Bijl-Feynman formula, provides a reasonable qualitative and semi-quantitative account of the elementary excitation spectrum in superfluid $^4$He, in particular of the presence and position of the roton minimum \cite{glyde}. 
\\ \indent
The behavior of the excitation spectrum as a function of $\sigma$ is the same across all values of density investigated in our simulations. Fig \ref{eq} shows two examples, for  $\rho=1.0$ and $\rho=100$. The phonon-like dispersion that is characteristic of the superfluid phase at low $q$ starts acquiring a negative curvature as the superfluid-supersolid transition is approached from above (with reference to the phase diagram shown in Fig. \ref{phase}a. At values of $\sigma$ sufficiently close to the critical value, the system experiences stronger density fluctuations at that value of the momentum, giving rise to a roton minimum and signaling incipient crystallization.
\\ \indent
These results are at least in qualitative agreement with the experimental data in Ref. \cite{ferlainopreprint}. As the value of $\sigma$ is lowered, the roton minimum becomes progressively lower, eventually hitting the horizontal axis; that is consistent with the divergence of $S(q)$ at the roton wave vector, i.e., the formation of a dipolar crystal. We come back to a more extensive comparison of our results with experiment in the next section.
\\ \indent
It is important to note that the presence of a roton minimum in the elementary excitation spectrum merely signals the proximity of the system to crystallization. As such, it is not a special feature of this particular system, or of the character of the interaction (dipolar).
Indeed, a roton minimum can be observed in the large $\sigma$ limit, in which the system is essentially a gas of hard spheres, as the crystal (in this case a conventional one, i.e., with few particles per unit cell) is approached from below. In the latter case, the roton minimum forms at a different, higher value of the wave vector, which reflects the fact that the ensuing crystal structure in this case has a smaller lattice constant.

\section{Conclusions}\label{conc}
We carried out extensive QMC simulations of a system of dipolar Bose particles of spin zero, in three dimensions, in order to gain insight into the phase diagram of this system. The anisotropic character of the interaction  gives rise to novel phases, chiefly a supersolid characterized by various, intriguing density patterns.
\\ \indent
The supersolid phase appears to be observable in a relatively broad range of parameter space, obviously making allowance for the difficulty of exploring the high density phase, as three-body recombination must be overcome.
\\ \indent
The results presented here are consistent with recent experimental data \cite{ferlainopreprint,tanzi,pfaupreprint,newferlaino}. In order to assess whether the density range explored here is comparable to that of recent experiments, we consider  Ref. \cite{newferlaino}, reporting measurements carried out on an assembly of $^{164}$Dy atoms. Using their numbers, namely $N=3.5\times{10^4}$ atoms confined in an anisotropic harmonic potential of characteristic frequencies equal to 300, 16 and 22 Hz, we estimate the density at the center of the trap  $\sim 1.3\times 10^{-8}$\ \AA$^{-3}$. Expressed in units of the dipolar length $a$, which in this case is worth 208 \AA\ \cite{factorofthree}, this is equal to 0.117, i.e., very close to the lowest density considered in this work. Indeed, the results shown in Fig. 1g of ref. \cite{newferlaino} indicate that the transition between a uniform BEC and a supersolid occurs in correspondence of a scattering length $a_s \sim 0.24\ a$, which is in remarkable quantitative agreement with our prediction of $\sigma\sim 0.4\ a$ for $\rho=0.125$ (Eq. \ref{as2} yields  $\sigma=0.38$ if $a_s=0.24$). 
\\ \indent
This result provides strong quantitative support for the microscopic model utilized here, as well as for the results of our calculation. It also suggests that the experimental findings of Ref. \cite{newferlaino} reflect the physics of the bulk, to an appreciable extent, i.e., they are not overly affected by confinement, nor by the relatively small size of the system.
This observation is consistent with the conclusions of a recent theoretical (mean-field) study of the system confined in an elongated trap \cite{ancilotto}, whose results are in qualitative agreement with ours. 

\section*{Acknowledgments}
This work was supported by the Natural Sciences and Engineering Research Council of Canada, as well as ComputeCanada. Useful conversations with F. Cinti, S. Moroni and F. Ferlaino are also gratefully acknowledged.

\section*{Conflict of interest}
The authors declare that they have no conflict of interest.


\begin{thebibliography}{99}
\bibitem{lastnote}
In a typical experiment the assembly of dipolar particles is spatially confined by means of a harmonic trap, which is generally non-spherical. Therefore, experimental observations will be to a degree affected by the specific confinement. However, the confining lengths in the three directions are generally large enough that one can reasonably argue that one is mostly investigating bulk properties.
\bibitem{morepr}
More precisely, Fig. \ref{str1}a, like other similar images featured in this paper, shows the particle density map (integrated over the $z$ direction) obtained from a statistically representative configuration (i.e., particle world lines). By “statistically” representative, we mean that every configuration generated in the
simulation is physically equivalent to that shown in the figure, differing at the most by a rotation and/or a translation.
\bibitem{super}
The superfluid properties of a single filament are of interest in their own right, and they will be the subject of future work. A regime of independent (quasi)superfluid filaments in the thermodynamic limit seems possible.
\bibitem{factorofthree}
It need be noted that the dipolar length defined in Ref. \cite{newferlaino} is equivalent to $1/3$ of that defined here.
\end{thebibliography}
\end{document}